\documentclass[conference]{IEEEtran}

\usepackage{cite}
\usepackage{amsmath,amssymb,amsfonts}
\usepackage{graphicx}
\usepackage{textcomp}
\usepackage{xcolor}
\usepackage{tabularray}
\UseTblrLibrary{booktabs}
\usepackage{enumerate}
\usepackage{listings}
\usepackage{url}
\usepackage{hyperref}
\usepackage{tikz}
\usetikzlibrary{positioning,calc,fit,decorations.pathreplacing}
\usepackage{pgfplots}
\pgfplotsset{compat=1.18}

\definecolor{figInputFill}{RGB}{218,213,204}
\definecolor{figAgentFill}{RGB}{178,194,218}
\definecolor{figOutputFill}{RGB}{198,213,228}
\definecolor{figEvalFill}{RGB}{212,208,222}
\definecolor{figBarFill}{RGB}{98,138,183}
\definecolor{figInputBorder}{RGB}{170,162,148}
\definecolor{figAgentBorder}{RGB}{110,135,170}
\definecolor{figOutputBorder}{RGB}{130,158,182}
\definecolor{figEvalBorder}{RGB}{148,140,168}
\definecolor{tblHeaderTint}{RGB}{220,228,240}   
\definecolor{tblRowAlt}{RGB}{246,246,248}        
\definecolor{tblAvgRow}{RGB}{238,238,238}        
\definecolor{tblSpreadRow}{RGB}{225,232,244}     
\definecolor{tblAmber}{RGB}{255,244,204}         
\definecolor{demoHigh}{RGB}{213,240,220}         
\definecolor{demoMedHigh}{RGB}{232,245,233}      
\definecolor{demoMid}{RGB}{255,249,196}          
\definecolor{demoMedLow}{RGB}{255,236,179}       
\definecolor{demoLow}{RGB}{255,218,185}          
\definecolor{demoNone}{RGB}{255,205,210}         

\begin{document}

\title{Architecture as Capability Equalizer\\for Coding Agents}

\author{
\IEEEauthorblockN{Arquimedes Canedo}
\IEEEauthorblockA{Siemens Digital Industries Software\\
Princeton, NJ, USA}
}

\maketitle

\begin{abstract}
LLM-based coding agents generate complete software systems from high-level descriptions, yet little is known about how the \emph{format} of architecture specifications affects the quality of generated code or whether this effect depends on model capability. We present a controlled experiment comparing five informationally equivalent specification formats (informal prose, Mermaid diagrams with constraints and ADRs, OpenAPI, C4/Structurizr DSL, and TypeScript interface contracts with ArchUnit-style rules) across six models from three vendor families (Anthropic Claude, OpenAI GPT, Google Gemini). Across 90 multi-turn agent trials, specification format shows a strong \emph{format $\times$ model interaction}. On the strongest models (Sonnet 4.6, GPT-5), format barely matters (quality spread $0.17$--$0.92$). On weaker models, format produces spreads of $0.83$--$2.42$ points, with code-proximate formats (OpenAPI, TypeScript contracts) recovering most of the capability gap. Mid-tier models can consume \emph{more} tokens than frontier models for worse output when they enter compilation debugging loops that stronger models avoid. Self-validation rates collapse from $100\%$ (Sonnet) to $0\%$ (Gemini Flash) across the capability spectrum. TypeScript contracts triple API route coverage for the weakest model ($33\% \rightarrow 100\%$). Structured architecture specifications serve as a \emph{capability equalizer}, with value inversely proportional to model strength and the largest returns for cost-optimized deployments.
\end{abstract}


\section{Introduction}

LLM-based coding agents (tools that iteratively write, compile, and debug code through multi-turn interactions) have shifted software engineering workflows from manual implementation toward specification and review~\cite{jimenez2024swebench}. Agents such as Claude Code, GitHub Copilot Workspace, and Cursor now generate entire modules from high-level descriptions. The question of how developers should communicate architectural intent to these agents remains open.

Current practice is ad hoc. Developers include an \texttt{ARCHITECTURE.md} file, embed guidance in system prompts, or rely on the agent to infer structure from existing code. When architecture guidance is provided, it typically takes the form of informal prose. This mirrors how human developers consume design documents but is not necessarily optimal for LLM consumption.

This paper starts from the premise that providing architecture specifications to coding agents is beneficial. We do not ask \emph{whether} architecture guidance helps, but rather \emph{how the format of that guidance} affects the outcome. (A baseline experiment without architecture guidance validates this premise; see \S\ref{sec:discussion}.) We hypothesize that more structured specification formats, those that separate structural description, behavioral constraints, and design rationale into distinct, machine-parseable sections, improve the quality of agent-generated code. To test this, we compare five specification formats across three quality dimensions:

\begin{enumerate}
    \item Architectural adherence: do component boundaries and communication patterns match the specification?
    \item Constraint compliance: are explicitly stated rules respected?
    \item Completeness: are all specified components and operations implemented?
\end{enumerate}

Our contributions are:
\begin{itemize}
    \item Format $\times$ model interaction. On the two frontier models (Sonnet, GPT-5), format barely matters (quality spread $0.17$--$0.92$). On non-frontier models, format produces spreads of $0.83$--$2.42$ points, with code-proximate formats recovering over half the capability gap.
    \item Inverted cost efficiency. Mid-tier models capable of iterative repair but not capable enough to succeed quickly can consume \emph{more} tokens than frontier models for worse output (Haiku: $735$K tokens, score $6.50$; Sonnet: $640$K tokens, score $8.42$), a ``valley'' in the capability--cost curve.
    \item Failure mode taxonomy. Three distinct agent failure modes (compilation death spiral, premature termination, and perfectionist iteration) each interact differently with specification format.
    \item Self-validation collapse. Demo run rate drops from $100\%$ (frontier) to $0\%$ (smallest model). Weaker agents never test their own output end-to-end, a concrete deployment risk invisible in benchmark scores.
    \item TypeScript contracts close the gap for the weakest model. Gemini Flash implements only $33\%$ of specified API routes under prose, but $100\%$ under TypeScript interface contracts, a $3\times$ improvement from specification format alone, measured by automated route coverage analysis.
    \item Hybrid evaluation methodology. Automated static constraint checking ($\sim$80\% architecture coverage) and LLM-as-judge scoring are complementary. The judge catches semantic violations the checker misses ($r = 0.21$ between judge constraint score and automated compliance), and vice versa.
    \item Open dataset. We release all 93 trial transcripts, generated codebases, judge scores, automated compliance results, and the experiment harness at \url{https://github.com/arquicanedo/architecture-as-equalizer}.
\end{itemize}

\section{Related Work}
\label{sec:related}

\subsection{Specification-Driven LLM Code Generation}

The closest prior work is BaxBench~\cite{vero2025baxbench}, which compared OpenAPI specifications against prose descriptions for backend code generation across 28 scenarios, 14 frameworks, and 3 LLMs, finding statistically significant gains of $+5.8\%$ to $+9.6\%$ pass@1 with OpenAPI. However, BaxBench uses single-turn generation with functional test evaluation and does not test architecture-level specifications, multi-turn agents, or architectural quality metrics. We extend BaxBench's finding from API-level functional specifications to architecture-level specifications (module decomposition, dependency rules, runtime invariants) in multi-turn agentic generation.

Dente et al.'s Constraint Decay~\cite{dente2026constraint} shows that adding architectural constraints \emph{on top of} a fixed OpenAPI specification degrades performance by an average of 30 percentage points across 80 tasks, 7 models, and 2 agent scaffolds, challenging the premise that more architectural structure always helps. Constraint Decay holds format constant and varies constraint \emph{density}. We hold content constant and vary \emph{format}, a complementary experimental design.

CodeSpec~\cite{wang2026codespec} shows that \emph{executable} architecture specifications (LLM-generated checkers for unit existence, relation preservation, and data flow) achieve $71.8\%$ pass rate versus $43.8\%$ for textual specifications on complex tasks. Their specifications provide iterative feedback during generation, while ours are human-authored declarative documents provided as static context. ConCodeEval~\cite{kammakomati2024concodeeval} directly compares five schema formats (JSON, YAML, XML, Python, natural language) for data-level constraint adherence and finds that format matters significantly. This is the closest precedent for our format comparison, though at the data-schema rather than architecture level. Shafin et al.~\cite{shafin2025process} find that imposing structured processes (Waterfall) on multi-agent class-level generation \emph{reduces} correctness for 2 of 3 models, cautioning that more structure does not always help.

\subsection{Structured Prompting}

The effect of prompt structure on LLM output quality is well established. Wei et al.\ showed that chain-of-thought prompting improves reasoning~\cite{wei2022chain}, while Jiang et al.\ found that structured task decomposition improves multi-file code generation~\cite{jiang2024selfplanning}. These studies focus on task-level prompting rather than architecture-level specification.

\subsection{Architecture Conformance}

Architectural erosion (the divergence of implementation from intended architecture) is well studied~\cite{perry1992foundations, terra2012qualitas}. Tools such as ArchUnit~\cite{archunit} enforce architectural rules through automated tests. Bogner et al.~\cite{bogner2024llmarchitecture} found that LLMs produce tightly coupled code without explicit architecture guidance. Konrad et al.~\cite{konrad2026architecture} propose a three-layer governance framework including fitness-function-style post-generation checks comparing dependency graphs against declared constraints, the conceptual framework we implement empirically. Meawad~\cite{meawad2026designfirst} proposes design-first governance using contract-driven constraint layers, arguing that governance structure matters more than model capability (only the abstract is accessible due to IEEE paywall). Our work extends this line by measuring whether the \emph{format} of architecture guidance affects conformance rates, and how this interacts with model capability.

\subsection{Diagrams-as-Code and API Specifications}

Text-based architecture formats (Mermaid~\cite{mermaid}, PlantUML~\cite{plantuml}, the C4 model~\cite{brown2018c4}) are widely adopted for version-controllable documentation. OpenAPI~\cite{openapi} provides formal, machine-readable API contracts. These formats are naturally parseable by LLMs, but no prior work has compared their effectiveness as architecture specification input for code generation agents. Our experiment is the first to test Mermaid diagrams, C4/Structurizr DSL, and TypeScript interface contracts as LLM input for system-level code generation.

\subsection{Multi-Turn Agents and Evaluation}

Yang et al.'s SWE-agent~\cite{yang2024sweagent} demonstrated that tool use significantly improves agent capability over single-turn generation. Jimenez et al.'s SWE-bench~\cite{jimenez2024swebench} evaluates agents on real GitHub issues but measures only functional correctness rather than architectural quality. For evaluation methodology, Zheng et al.~\cite{zheng2023judging} established LLM-as-judge with $85\%$ agreement with human experts, though self-preference bias is documented for same-model evaluation. Vasilevski et al.~\cite{vasilevski2026beyondcorrectness} validated LLM-as-judge specifically for \emph{architectural quality} of generated code, using repository-derived rubrics with categorical verdicts. This is the closest precedent to our architectural judge, though they evaluate patches to existing code rather than greenfield generation. Our harness adopts the multi-turn paradigm with a hybrid evaluation combining automated static constraint checking and LLM-as-judge scoring.

\subsection{Cross-Model Capability Effects}

Prior benchmarks have documented large performance gaps between model tiers on code generation tasks: on SWE-bench, frontier models resolve $30$--$50\%$ of issues while smaller models resolve under $5\%$~\cite{jimenez2024swebench}. However, no study has measured whether input specification format \emph{moderates} this capability gap, i.e., whether structured specifications help smaller models disproportionately. Our cross-model experiment (Claude Sonnet 4.6~\cite{anthropic2026claude} vs.\ Claude Haiku 4.5) directly tests this interaction, finding that specification format effects are $10\times$ larger on the smaller model.

\section{Approach}
\label{sec:approach}

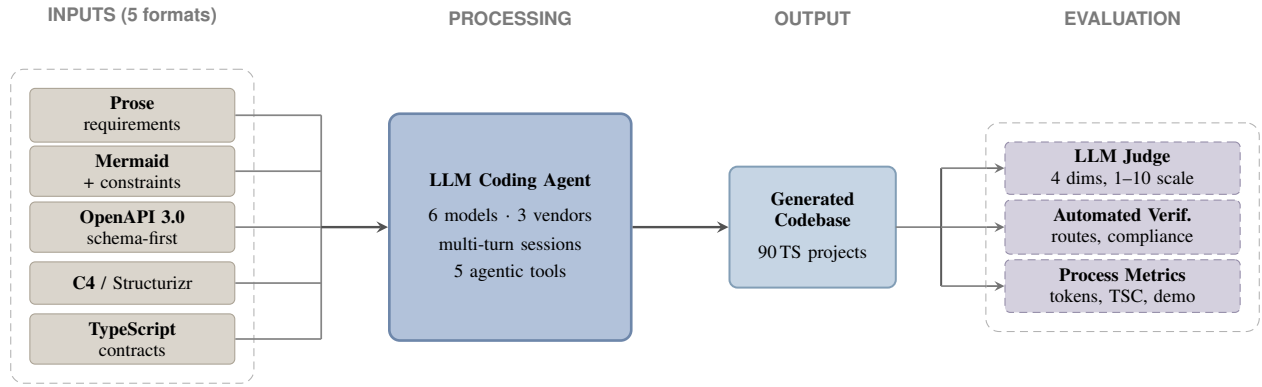
\begin{figure*}[htbp]
\centering
\begin{tikzpicture}[
  every node/.style={font=\scriptsize},
  specbox/.style={
    draw=figInputBorder, rounded corners=2pt,
    minimum height=0.56cm, text width=2.50cm,
    align=center, fill=figInputFill, inner sep=3pt
  },
  agentbox/.style={
    draw=figAgentBorder, rounded corners=4pt, line width=1.0pt,
    minimum width=3.20cm, minimum height=3.00cm,
    align=center, fill=figAgentFill, inner sep=6pt
  },
  outbox/.style={
    draw=figOutputBorder, rounded corners=3pt, line width=0.75pt,
    minimum width=2.20cm, minimum height=1.60cm,
    align=center, fill=figOutputFill, inner sep=4pt
  },
  evalbox/.style={
    draw=figEvalBorder, rounded corners=2pt, densely dashed,
    minimum height=0.64cm, text width=2.90cm,
    align=center, fill=figEvalFill, inner sep=3pt
  },
  groupbox/.style={draw=black!30, rounded corners=5pt,
                   densely dashed, inner sep=7pt},
  mainarr/.style={->, >=stealth, line width=0.9pt, black!68},
  thinarr/.style={->, >=stealth, line width=0.65pt, black!55},
  busline/.style={line width=0.65pt, black!48},
  slab/.style={font=\scriptsize\bfseries\sffamily,
               text=black!52, anchor=south},
]

\node[minimum size=0pt, inner sep=0pt] at (0.00, 0) {};
\node[minimum size=0pt, inner sep=0pt] at (16.20, 0) {};

\node[specbox] at (1.50, 1.30) (prose)   {\textbf{Prose}\\requirements};
\node[specbox] at (1.50, 0.56) (mermaid) {\textbf{Mermaid}\\+ constraints};
\node[specbox] at (1.50,-0.18) (openapi) {\textbf{OpenAPI 3.0}\\schema-first};
\node[specbox] at (1.50,-0.92) (c4)      {\textbf{C4} / Structurizr};
\node[specbox] at (1.50,-1.66) (ts)      {\textbf{TypeScript}\\contracts};
\node[groupbox, fit=(prose)(ts)] (specgrp) {};

\node[agentbox] at (6.50,-0.18) (agent) {%
  \textbf{LLM Coding Agent}\\[5pt]%
  6 models $\cdot$ 3 vendors\\[3pt]%
  multi-turn sessions\\[2pt]%
  5 agentic tools%
};

\node[outbox] at (10.50,-0.18) (code) {%
  \textbf{Generated}\\%
  \textbf{Codebase}\\[4pt]%
  90\,TS projects%
};

\node[evalbox] at (14.60, 0.60) (judge)
      {\textbf{LLM Judge}\\4 dims, 1--10 scale};
\node[evalbox] at (14.60,-0.18) (autov)
      {\textbf{Automated Verif.}\\routes, compliance};
\node[evalbox] at (14.60,-0.96) (proc)
      {\textbf{Process Metrics}\\tokens, TSC, demo};
\node[groupbox, fit=(judge)(proc)] (evalgrp) {};

\coordinate (LY) at ($(specgrp.north) + (0, 0.45cm)$);
\node[slab] at (LY -| specgrp) {INPUTS (5 formats)};
\node[slab] at (LY -| agent)   {PROCESSING};
\node[slab] at (LY -| code)    {OUTPUT};
\node[slab] at (LY -| evalgrp) {EVALUATION};

\coordinate (tx) at ($(specgrp.east)!0.50!(agent.west)$);
\foreach \s in {prose,mermaid,openapi,c4,ts}{
  \draw[busline] (\s.east) -- (tx |- \s.east);
}
\draw[busline] (tx |- prose.east) -- (tx |- ts.east);
\draw[mainarr] (tx |- agent.west) -- (agent.west);
\draw[mainarr] (agent.east) -- (code.west);

\draw[busline] (11.60,-0.18) -- (12.20,-0.18);
\draw[line width=0.75pt, black!65] (12.20, 0.60) -- (12.20,-0.96);
\draw[busline] (12.20, 0.60) -- (12.80, 0.60);
\draw[busline] (12.20,-0.18) -- (12.80,-0.18);
\draw[busline] (12.20,-0.96) -- (12.80,-0.96);
\draw[thinarr] (12.80, 0.60) -- (judge.west);
\draw[thinarr] (12.80,-0.18) -- (autov.west);
\draw[thinarr] (12.80,-0.96) -- (proc.west);

\end{tikzpicture}
\caption{Experimental pipeline. Five architecture specification formats are provided to 6 LLMs across 3 vendor families. Each multi-turn agent session produces a TypeScript codebase evaluated through three independent channels: an LLM judge (4 dimensions, 1--10 scale), automated verification (route coverage, constraint compliance), and process metrics (token cost, compilation, demo run). A separate no-architecture baseline validates the premise (\S\ref{sec:discussion}).}
\label{fig:pipeline}
\end{figure*}

Figure~\ref{fig:pipeline} summarizes the experimental pipeline. We design a reference system, describe its architecture in five informationally equivalent formats, give each to six LLMs from three vendor families, and evaluate the generated code through three independent channels.

\subsection{System Under Test}

We designed a Task Management API with sufficient architectural complexity to expose meaningful differences between specification formats. The system comprises seven components:

\begin{itemize}
    \item API Router: HTTP entry point using Node.js built-in \texttt{http} module
    \item User Service: User CRUD operations
    \item Project Service: Project CRUD with member management
    \item Task Service: Task CRUD with status transitions (\texttt{todo} $\rightarrow$ \texttt{in-progress} $\rightarrow$ \texttt{done})
    \item Comment Service: Comments on tasks
    \item Notification Service: Event-driven notification creation
    \item Event Bus: In-memory publish/subscribe for inter-service communication
\end{itemize}

The system uses exclusively in-memory storage with no external dependencies, making it fully self-contained. The architectural complexity arises from inter-service communication patterns, data ownership rules, and the constraint that services must communicate through the event bus rather than direct calls.

\subsection{What We Mean by ``Architecture''}

In this experiment, ``architecture'' is a complete system-level specification comprising seven elements:

\begin{enumerate}
    \item Components. Seven building blocks: an API Router, five domain services (User, Project, Task, Comment, Notification), and an Event Bus.
    \item Communication patterns. Services never call each other directly. All inter-service communication goes through the Event Bus via publish/subscribe on named events (\texttt{task.assigned}, \texttt{task.statusChanged}, \texttt{comment.added}).
    \item Data ownership. Each service exclusively owns its own in-memory data store. No service reads or writes another's store.
    \item Behavioral constraints. Six hard rules: no direct service-to-service calls, exclusive data ownership, HTTP handling only in the Router, forward-only status transitions (\texttt{todo} $\rightarrow$ \texttt{in-progress} $\rightarrow$ \texttt{done}), no npm dependencies, one service per file.
    \item API surface. 25 HTTP routes mapping methods and paths to service operations.
    \item Design rationale. Three architectural decision records explaining \emph{why}: event bus for decoupling, service-owned stores to prevent shared-state bugs, no frameworks for self-containment.
    \item File structure. A prescribed directory layout (\texttt{src/event-bus.ts}, \texttt{src/services/*.ts}, \texttt{src/router.ts}, etc.).
\end{enumerate}

All five specification formats encode the same architecture (same components, same constraints, same routes, same rationale). Only the \emph{representation} differs. The experiment measures whether the representation format affects how well agents implement the architecture.

\subsection{Specification Formats}

We created five informationally equivalent specifications. All contain the same components, operations, constraints, communication patterns, and design rationale. Only the format differs.

\subsubsection{Prose (Control)}
A natural language document describing the system in flowing paragraphs. This mirrors how a senior developer might describe a system architecture in a design document.

\subsubsection{Mermaid + Constraints + ADRs}
A specification combining Mermaid architecture and sequence diagrams, structured per-service component blocks, a numbered constraint list, architectural decision records (ADRs), and a route table.

\subsubsection{OpenAPI + Mermaid + Constraints}
Extends the Mermaid+constraints format with a complete OpenAPI 3.0 specification defining all routes, request/response schemas, status codes, and data models in YAML. This provides a formal, machine-checkable API contract.

\subsubsection{C4/Structurizr DSL}
Uses the C4 model's hierarchical decomposition (System Context $\rightarrow$ Container $\rightarrow$ Component) expressed in Structurizr DSL, supplemented with a Mermaid container diagram, component detail tables, and the same constraints and ADRs.

\subsubsection{TypeScript Contracts + Architecture Rules}
Provides the exact TypeScript interfaces for all data models and service contracts (\texttt{IUserService}, \texttt{ITaskService}, etc.), event payload types, and ArchUnit-style architecture rules expressed as pseudo-code declarations (e.g., \texttt{RULE: Files matching ``services/*-service.ts'' MUST NOT import from other service files}).

\subsubsection{Format Comparison}

Table~\ref{tab:format_comparison} illustrates how the same architectural constraint --- that services must not call each other directly --- is expressed in each format. The progression from natural language to typed rules shows the increasing level of machine-parseable specificity.

\begin{table*}[htbp]
\centering
\caption{The Same Architecture in Five Formats. Each excerpt shows how inter-service communication is specified. All encode identical constraints; only the representation differs.}
\label{tab:format_comparison}
\begin{tblr}{
  colspec = {l X},
  width = \textwidth,
  rowsep = 4pt,
  row{1} = {font=\bfseries, bg=tblHeaderTint},
  row{3,5} = {bg=tblRowAlt},
  column{1} = {wd=2.5cm},
  column{2} = {font=\small\ttfamily},
  hline{1,Z} = {0.8pt},
  hline{2} = {0.5pt},
  hline{3,4,5,6} = {0.3pt, dashed},
}
{Format} & {Specification Excerpt} \\
{Prose} & {\textnormal{\rmfamily Services should not call each other directly. Instead, we use a simple in-memory Event Bus --- a publish/subscribe system. When something notable happens, the originating service publishes an event to the Event Bus. Other services that care about those events subscribe to them and react accordingly.}} \\
{Mermaid + Constr.} & {\textcolor{blue!70!black}{graph} TD \newline \hspace*{1em}TaskSvc \textcolor{red!70!black}{-->|publish|} EventBus[\textcolor{teal}{Event Bus}] \newline \hspace*{1em}CommentSvc \textcolor{red!70!black}{-->|publish|} EventBus \newline \hspace*{1em}EventBus \textcolor{red!70!black}{-->|subscribe|} NotifSvc \newline \hspace*{1em}UserSvc --- UserStore[(\textcolor{teal}{User Store})] \newline \textcolor{gray}{\% Constraint 1: Services MUST NOT import or call} \newline \textcolor{gray}{\% other services directly.}} \\
{OpenAPI 3.0} & {\textcolor{blue!70!black}{paths}: \newline \hspace*{1em}\textcolor{orange!80!black}{/tasks/\{id\}/assign}: \newline \hspace*{2em}\textcolor{blue!70!black}{put}: \newline \hspace*{3em}\textcolor{blue!70!black}{summary}: \textcolor{teal}{Assign task to user} \newline \hspace*{3em}\textcolor{blue!70!black}{requestBody}: \newline \hspace*{4em}\textcolor{blue!70!black}{schema}: \newline \hspace*{5em}\textcolor{blue!70!black}{properties}: \newline \hspace*{6em}\textcolor{blue!70!black}{assigneeId}: \{ \textcolor{blue!70!black}{type}: string \} \newline \hspace*{3em}\textcolor{blue!70!black}{responses}: \newline \hspace*{4em}\textcolor{orange!80!black}{'200'}: \{ \textcolor{blue!70!black}{\$ref}: \textcolor{teal}{Task} \}} \\
{C4 / Structurizr} & {\textcolor{blue!70!black}{workspace} \{ \newline \hspace*{1em}\textcolor{blue!70!black}{model} \{ \newline \hspace*{2em}taskService = \textcolor{blue!70!black}{container} \textcolor{teal}{"Task Service"} \{ \newline \hspace*{3em}taskStore = \textcolor{blue!70!black}{component} \textcolor{teal}{"Task Store"} \textcolor{gray}{"Map<string,Task>"} \newline \hspace*{3em}taskOps = \textcolor{blue!70!black}{component} \textcolor{teal}{"Task Operations"} \} \newline \hspace*{2em}taskService \textcolor{red!70!black}{->} eventBus \textcolor{teal}{"Publishes task.assigned"} \newline \hspace*{2em}commentService \textcolor{red!70!black}{->} eventBus \textcolor{teal}{"Publishes comment.added"} \newline \hspace*{2em}eventBus \textcolor{red!70!black}{->} notificationService \textcolor{teal}{"Delivers events to"} \} \}} \\
{TS Contracts} & {\textcolor{blue!70!black}{interface} \textcolor{teal}{ITaskService} \{ \newline \hspace*{1em}create(input: \textcolor{teal}{CreateTaskInput}): \textcolor{teal}{Task}; \newline \hspace*{1em}assign(taskId: \textcolor{teal}{string}, assigneeId: \textcolor{teal}{string}): \textcolor{teal}{Task}; \newline \hspace*{1em}changeStatus(taskId: \textcolor{teal}{string}, newStatus: \textcolor{teal}{TaskStatus}): \textcolor{teal}{Task}; \} \newline \newline \textcolor{blue!70!black}{RULE 1}: NO\_CROSS\_SERVICE\_IMPORTS \newline \hspace*{1em}Files matching \textcolor{teal}{"services/*-service.ts"} \newline \hspace*{1em}\textcolor{red!70!black}{MUST NOT} import from other \textcolor{teal}{"services/*-service.ts"}} \\
\end{tblr}
\end{table*}

\subsection{Experimental Harness}

We built a TypeScript-based experiment runner that creates independent LLM agent sessions for each trial. The harness provides the agent with five tools: \texttt{write\_file}, \texttt{read\_file}, \texttt{list\_files}, \texttt{run\_command}, and \texttt{done}. Each trial proceeds as a multi-turn conversation in which the agent receives the specification, then iteratively writes files, compiles, fixes errors, runs demos, and signals completion.

\subsection{Evaluation}

Automated analysis checks for constraint violations via static import analysis, structural completeness, and TypeScript compilation.

LLM judge evaluation scores each trial's output (blind to specification type) on four dimensions using a 1--10 scale, covering architectural adherence, completeness, code quality, and constraint compliance~\cite{zheng2023judging}.

\section{Experimental Setup}
\label{sec:setup}

We ran six complete rounds of the experiment, one per model, to test format $\times$ model $\times$ vendor interactions.

\begin{itemize}
    \item Anthropic: Claude Sonnet 4.6 (frontier), Claude Haiku 4.5 (mid-tier)
    \item OpenAI: GPT-5 (frontier), GPT-5-mini (mid-tier)
    \item Google: Gemini 2.5 Pro (mid-tier), Gemini 2.5 Flash (small)
    \item Total: 90 trials across 30 cells (5 formats $\times$ 6 models)
\end{itemize}

Shared parameters across all rounds:
\begin{itemize}
    \item Max output tokens per turn: 16{,}384
    \item Max turns per trial: 50
    \item Temperature: Default
    \item Tools: \texttt{write\_file}, \texttt{read\_file}, \texttt{list\_files}, \texttt{run\_command}, \texttt{done}
    \item Judge model: Claude Sonnet 4.6 for all rounds (ensuring consistent evaluation)
\end{itemize}

Each trial starts from a clean directory with no prior context. The implementation prompt is identical across all conditions and models. Using the same judge model across all rounds ensures that score differences reflect code quality rather than judge capability.

\section{Results}
\label{sec:results}

\subsection{Sonnet 4.6 Results (Frontier Model)}

Table~\ref{tab:master} presents all 15 Sonnet trials. Table~\ref{tab:summary} aggregates by condition.

\definecolor{bestcell}{HTML}{E8F5E9}
\definecolor{worstcell}{HTML}{FFEBEE}
\definecolor{groupheader}{RGB}{225,232,244}

\begin{table*}[htbp]
\centering
\caption{Sonnet 4.6 --- Complete Trial Results: Judge Scores (1--10), Resource Consumption, and Constraint Violations for All 15 Trials. \colorbox{bestcell}{\strut}\,Best trial, \colorbox{worstcell}{\strut}\,Worst trial. Arch = Architectural Adherence, Comp = Completeness, Qual = Code Quality, Constr = Constraint Compliance, LoC = Lines of Code, Errors = Tool call errors during agent session.}
\label{tab:master}
\begin{tblr}{
  colspec = {l l c c c c c r r r r c},
  width = \textwidth,
  rowsep = 2pt,
  column{1} = {wd=2.2cm},
  row{1} = {font=\bfseries\small, bg=tblHeaderTint},
  row{2} = {font=\bfseries, bg=tblHeaderTint},
  row{4,7,10,13,16} = {bg=tblRowAlt},
  hline{1,Z} = {0.8pt},
  hline{2} = {3-6}{0.4pt},
  hline{2} = {8-11}{0.4pt},
  hline{3} = {0.5pt},
  hline{6,9,12,15} = {0.3pt, dashed},
  cell{7}{7} = {bg=bestcell},
  cell{8}{7} = {bg=worstcell},
  cell{13}{7} = {bg=bestcell},
}
 & & \SetCell[c=4]{c} \textsc{Judge Scores} & & & & & \SetCell[c=4]{c} \textsc{Resource Consumption} & & & & \textsc{Constr.} \\
{Condition} & {Trial} & {Arch} & {Comp} & {Qual} & {Constr} & {Overall} & {Tokens (K)} & {Turns} & {LoC} & {Errors} & {Violations} \\
\SetCell[r=3]{l} {Prose} & T1 & 8 & 9 & 9 & 8 & 8.50 & 367 & 23 & 1{,}823 & 2 & 0 \\
 & T2 & 8 & 9 & 9 & 8 & 8.50 & 602 & 28 & 1{,}810 & 3 & 0 \\
 & T3 & 7 & 9 & 9 & 8 & 8.25 & 332 & 24 & 1{,}424 & 2 & 0 \\
\SetCell[r=3]{l} {Mermaid + Constr. + ADRs} & T1 & 9 & 10 & 9 & 10 & \textbf{9.50} & 1{,}735 & 42 & 1{,}576 & 4 & 1 \\
 & T2 & 8 & 9 & 9 & 9 & 8.75 & 412 & 24 & 1{,}332 & 2 & 0 \\
 & T3 & 6 & 9 & 8 & 6 & 7.25 & 1{,}032 & 40 & 1{,}380 & 4 & 1 \\
\SetCell[r=3]{l} {OpenAPI + Mermaid + Constr.} & T1 & 7 & 9 & 9 & 7 & 8.00 & 522 & 25 & 1{,}573 & 2 & 0 \\
 & T2 & 9 & 9 & 9 & 8 & 8.75 & 334 & 20 & 1{,}522 & 2 & 0 \\
 & T3 & 8 & 9 & 9 & 8 & 8.50 & 517 & 27 & 1{,}539 & 4 & 0 \\
\SetCell[r=3]{l} {C4 / Structurizr DSL} & T1 & 9 & 10 & 9 & 10 & \textbf{9.50} & 880 & 32 & 1{,}776 & 4 & 0 \\
 & T2 & 7 & 9 & 8 & 7 & 7.75 & 518 & 22 & 1{,}492 & 2 & 2 \\
 & T3 & 7 & 9 & 8 & 7 & 7.75 & 502 & 23 & 1{,}380 & 4 & 0 \\
\SetCell[r=3]{l} {TS Contracts + ArchUnit Rules} & T1 & 8 & 9 & 9 & 8 & 8.50 & 960 & 32 & 1{,}392 & 5 & 0 \\
 & T2 & 7 & 9 & 9 & 7 & 8.00 & 456 & 28 & 1{,}286 & 5 & 0 \\
 & T3 & 8 & 9 & 9 & 9 & 8.75 & 429 & 24 & 1{,}254 & 5 & 0 \\
\end{tblr}
\end{table*}

\begin{table*}[htbp]
\centering
\caption{Sonnet 4.6 --- Aggregated Results by Condition (mean $\pm$ range across 3 trials; \textbf{bold} = best in row)}
\label{tab:summary}
\begin{tblr}{
  colspec = {X c c c c c},
  width = \textwidth,
  rowsep = 3pt,
  row{1} = {font=\bfseries, bg=tblHeaderTint},
  row{4,6,8} = {bg=tblRowAlt},
  row{11,13} = {bg=tblRowAlt},
  row{16} = {bg=tblRowAlt},
  hline{1,Z} = {0.8pt},
  hline{2} = {0.5pt},
  hline{3} = {0.3pt},
  hline{9,14,17} = {0.5pt},
  hline{10,15,18} = {0.3pt},
  cell{2}{1} = {c=6}{l,font=\itshape,bg=groupheader},
  cell{9}{1} = {c=6}{l,font=\itshape,bg=groupheader},
  cell{14}{1} = {c=6}{l,font=\itshape,bg=groupheader},
  cell{17}{1} = {c=6}{l,font=\itshape,bg=groupheader},
}
{Metric} & {Prose} & {Mermaid + Constr.} & {OpenAPI} & {C4} & {TS Contracts} \\
{Judge Scores (1--10 scale, higher is better)} & & & & & \\
Architectural Adherence & $7.67$ & $7.67$ & $\mathbf{8.00}$ & $7.67$ & $7.67$ \\
Completeness & $9.00$ & $\mathbf{9.33}$ & $9.00$ & $\mathbf{9.33}$ & $9.00$ \\
Code Quality & $\mathbf{9.00}$ & $8.67$ & $\mathbf{9.00}$ & $8.33$ & $\mathbf{9.00}$ \\
Constraint Compliance & $8.00$ & $\mathbf{8.33}$ & $7.67$ & $8.00$ & $8.00$ \\
\textbf{Overall} & $8.42$ & $\mathbf{8.50}$ & $8.42$ & $8.33$ & $8.42$ \\
Score Range & $\mathbf{0.25}$ & $2.25$ & $0.75$ & $1.75$ & $0.75$ \\
{Resource Consumption (mean)} & & & & & \\
Total Tokens (K) & $\mathbf{433}$ & $1{,}060$ & $458$ & $634$ & $615$ \\
Agent Turns & $23.7$ & $38.3$ & $\mathbf{23.0}$ & $28.7$ & $32.0$ \\
Lines of Code & $1{,}686$ & $1{,}429$ & $1{,}545$ & $1{,}549$ & $\mathbf{1{,}311}$ \\
Tool Call Errors & $\mathbf{2.0}$ & $4.0$ & $2.3$ & $2.7$ & $5.3$ \\
{Constraint Violations (automated static analysis)} & & & & & \\
Total Violations & $\mathbf{0}$ & $2$ & $\mathbf{0}$ & $2$ & $\mathbf{0}$ \\
Trials with Zero & $\mathbf{3/3}$ & $1/3$ & $\mathbf{3/3}$ & $2/3$ & $\mathbf{3/3}$ \\
{Cost--Quality Efficiency} & & & & & \\
Score per 100K tokens & $\mathbf{1.94}$ & $0.80$ & $1.84$ & $1.31$ & $1.37$ \\
\end{tblr}
\end{table*}

\subsection{Key Observations}

Overall scores are close. Means range from $8.33$ (C4) to $8.50$ (Mermaid+constraints), a spread of only $0.17$ points. No single format dominates across all dimensions.

Within-condition variance exceeds between-condition variance. The best and worst trials across the entire experiment both come from structured formats. Mermaid-1 and C4-1 scored $9.50$, while Mermaid-3 scored $7.25$. Prose has the tightest distribution (range $0.25$), indicating that structured formats raise the quality \emph{ceiling} while introducing volatility.

OpenAPI achieves zero constraint violations. Together with prose and TypeScript contracts, OpenAPI produced no automated constraint violations across any trial. The formats with the most explicit constraint lists (Mermaid, C4) did not achieve the best automated compliance.

Prose is the most cost-efficient. At $433$K tokens mean, prose costs $2.4\times$ less than Mermaid+constraints ($1{,}060$K) while delivering comparable overall scores ($8.42$ vs.\ $8.50$).

\subsection{Qualitative Analysis}

The judge's detailed notes reveal patterns that the aggregate scores obscure.

The comment enrichment problem. All conditions struggled with the same cross-cutting concern. When a comment is created, the notification needs the task's assignee ID, but the Comment Service cannot call the Task Service. The resolution quality varied.

\begin{itemize}
    \item \emph{Prose} trials used ad hoc workarounds: event interception, router-mediated re-publication, or silent constraint violation.
    \item \emph{Mermaid+constraints} top trials resolved it cleanly by having the router pass enrichment data at creation time, with code comments referencing the ADRs.
    \item \emph{OpenAPI} trials benefited from the formal route specification making the router's orchestration role unambiguous.
    \item \emph{C4} trials used the hierarchical decomposition to reason about which layer should handle enrichment.
    \item \emph{TypeScript contracts} trials were guided by the interface signatures, which made it explicit what data each service method accepts.
\end{itemize}

Folder structure adherence. All structured conditions more consistently placed services in a \texttt{src/services/} subdirectory matching their specifications. Prose trials varied between flat and nested structures.

High-scoring trials. Both $9.50$-scoring trials (Mermaid-1 and C4-1) shared common patterns: custom error classes, clean event bus isolation, and the judge noting that the agent appeared to ``follow the specification methodically.'' These trials also consumed the most tokens in their conditions, indicating that thorough specification adherence requires more iteration.

\subsection{Haiku 4.5 Results (Smaller Model)}

To test whether specification format interacts with model capability, we repeated the experiment with Claude Haiku 4.5, a smaller, faster, cheaper model, using the same judge (Sonnet 4.6) for consistent scoring. Table~\ref{tab:haiku_master} presents the complete Haiku results.

\begin{table*}[htbp]
\centering
\caption{Haiku 4.5 --- Complete Trial Results (judged by Sonnet 4.6; GPT-5 re-judging confirmed consistent rankings, $r = 0.60$, see \S VII-E)}
\label{tab:haiku_master}
\begin{tblr}{
  colspec = {l l c c c c c r r r c},
  width = \textwidth,
  rowsep = 2pt,
  column{1} = {wd=2.2cm},
  row{1} = {font=\bfseries\small, bg=tblHeaderTint},
  row{2} = {font=\bfseries, bg=tblHeaderTint},
  row{4,7,10,13,16} = {bg=tblRowAlt},
  hline{1,Z} = {0.8pt},
  hline{2} = {3-6}{0.4pt},
  hline{2} = {8-10}{0.4pt},
  hline{3} = {0.5pt},
  hline{6,9,12,15} = {0.3pt, dashed},
  cell{4}{7} = {bg=worstcell},
  cell{5}{7} = {bg=worstcell},
  cell{8}{7} = {bg=bestcell},
  cell{11}{7} = {bg=bestcell},
}
 & & \SetCell[c=4]{c} \textsc{Judge Scores} & & & & & \SetCell[c=3]{c} \textsc{Resources} & & & \textsc{Constr.} \\
{Condition} & {Trial} & {Arch} & {Comp} & {Qual} & {Constr} & {Overall} & {Tok.\ (K)} & {Turns} & {LoC} & {Viol.} \\
\SetCell[r=3]{l} {Prose} & T1 & 4 & 6 & 7 & 4 & 5.25 & 558 & 30 & 1{,}319 & 7 \\
 & T2 & 5 & 6 & 6 & 4 & 5.25 & 506 & 26 & 1{,}334 & 1 \\
 & T3 & 6 & 7 & 7 & 5 & 6.25 & 867 & 38 & 1{,}350 & 2 \\
\SetCell[r=3]{l} {Mermaid + Constr. + ADRs} & T1 & 5 & 7 & 7 & 4 & 5.75 & 535 & 29 & 1{,}618 & 2 \\
 & T2 & 8 & 7 & 8 & 9 & \textbf{8.00} & 573 & 30 & 1{,}398 & 0 \\
 & T3 & 5 & 6 & 7 & 4 & 5.50 & 708 & 35 & 1{,}579 & 2 \\
\SetCell[r=3]{l} {OpenAPI + Mermaid + Constr.} & T1 & 6 & 7 & 7 & 6 & 6.50 & 845 & 35 & 1{,}681 & 2 \\
 & T2 & 8 & 7 & 8 & 9 & \textbf{8.00} & 823 & 32 & 1{,}759 & 0 \\
 & T3 & 7 & 7 & 8 & 7 & 7.25 & 974 & 40 & 1{,}173 & 0 \\
\SetCell[r=3]{l} {C4 / Structurizr DSL} & T1 & 6 & 7 & 7 & 6 & 6.50 & 644 & 28 & 1{,}234 & 0 \\
 & T2 & 6 & 7 & 8 & 7 & 7.00 & 919 & 34 & 1{,}158 & 0 \\
 & T3 & 5 & 7 & 7 & 4 & 5.75 & 633 & 28 & 1{,}146 & 2 \\
\SetCell[r=3]{l} {TS Contracts + ArchUnit Rules} & T1 & 7 & 8 & 8 & 7 & 7.50 & 736 & 30 & 1{,}160 & 0 \\
 & T2 & 5 & 8 & 7 & 5 & 6.25 & 971 & 38 & 1{,}072 & 0 \\
 & T3 & 7 & 8 & 8 & 7 & 7.50 & 735 & 30 & 1{,}268 & 0 \\
\end{tblr}
\end{table*}

\subsection{Cross-Model Comparison}

Table~\ref{tab:crossmodel} presents the central finding. Specification format interacts strongly with model capability.

\begin{table*}[htbp]
\centering
\caption{Format $\times$ Model Interaction: Mean Overall Scores Across 6 Models and 3 Vendor Families (90 trials). Format spread = max $-$ min overall score within each model. Larger spread = format matters more. \textbf{Bold} = best format per model.}
\label{tab:crossmodel}
\begin{tblr}{
  colspec = {X c c c c c c},
  width = \textwidth,
  rowsep = 3pt,
  row{1} = {font=\bfseries\small, bg=tblHeaderTint},
  row{2} = {font=\bfseries, bg=tblHeaderTint},
  row{4,6} = {bg=tblRowAlt},
  row{8} = {font=\itshape, bg=tblAvgRow},
  row{9} = {font=\itshape\bfseries, bg=tblSpreadRow},
  hline{1,Z} = {0.8pt},
  hline{2} = {2-3}{0.4pt},
  hline{2} = {4-5}{0.4pt},
  hline{2} = {6-7}{0.4pt},
  hline{3} = {0.5pt},
  hline{8} = {0.3pt},
  hline{9} = {0.3pt},
}
 & \SetCell[c=2]{c} \textsc{Anthropic} & & \SetCell[c=2]{c} \textsc{OpenAI} & & \SetCell[c=2]{c} \textsc{Google} & \\
{Format} & {Sonnet 4.6} & {Haiku 4.5} & {GPT-5} & {GPT-5-mini} & {Gem.\ Pro} & {Gem.\ Flash} \\
Prose & $8.42$ & $5.58$ & $7.33$ & $6.08$ & $5.83$ & $5.75$ \\
Mermaid + Constr. & $\mathbf{8.50}$ & $6.42$ & $7.17$ & $\mathbf{6.92}$ & $6.42$ & $6.00$ \\
OpenAPI & $8.42$ & $\mathbf{7.25}$ & $6.58$ & $6.83$ & $\mathbf{6.92}$ & $6.50$ \\
C4 / Structurizr & $8.33$ & $6.42$ & $6.83$ & $6.75$ & $6.67$ & $\mathbf{6.83}$ \\
TS Contracts & $8.42$ & $7.08$ & $\mathbf{7.50}$ & $6.83$ & $4.50$ & $6.67$ \\
\textit{Average} & $8.42$ & $6.55$ & $7.08$ & $6.68$ & $6.07$ & $6.35$ \\
\textit{Format spread} & $0.17$ & $1.67$ & $0.92$ & $0.83$ & $\mathbf{2.42}$ & $1.08$ \\
\end{tblr}
\end{table*}

On Sonnet, the format spread is $0.17$ points, effectively noise. On smaller models, the spread ranges from $0.83$ (GPT-5-mini) to $2.42$ (Gemini Pro), up to $14\times$ larger. The best format varies by model. Mermaid+constraints leads on Sonnet, TypeScript contracts on GPT-5, OpenAPI on Haiku and Gemini Pro, and C4 on Gemini Flash. No single structured format dominates across all models, but all structured formats outperform prose on smaller models.

\begin{figure}[htbp]
\centering
\newlength{\cellw}\setlength{\cellw}{1.08cm}
\newlength{\cellh}\setlength{\cellh}{0.78cm}
\begin{tikzpicture}
\definecolor{s45}{RGB}{235,200,195}  
\definecolor{s55}{RGB}{228,213,204}  
\definecolor{s58}{RGB}{218,213,204}  
\definecolor{s60}{RGB}{208,213,214}  
\definecolor{s64}{RGB}{198,213,228}  
\definecolor{s66}{RGB}{188,204,222}  
\definecolor{s68}{RGB}{178,194,218}  
\definecolor{s69}{RGB}{162,182,210}  
\definecolor{s71}{RGB}{142,168,200}  
\definecolor{s73}{RGB}{122,152,190}  
\definecolor{s75}{RGB}{105,140,183}  
\definecolor{s84}{RGB}{78,115,162}   
\definecolor{s85}{RGB}{68,105,152}   
\newcommand{\hcell}[5]{%
  \fill[#4, rounded corners=2pt]
    (#1*\cellw+0.04cm, -#2*\cellh-0.04cm)
    rectangle +({\cellw-0.08cm}, {-\cellh+0.08cm});
  \node[font=\scriptsize\bfseries, text=#5]
    at (#1*\cellw+0.5*\cellw, -#2*\cellh-0.5*\cellh) {#3};
}
\foreach \lbl/\idx in {Prose/0, Merm.+C./1, OpenAPI/2, C4/3, TS~Contr./4} {
  \node[font=\scriptsize\sffamily, anchor=east, text=black!80]
    at (-0.12cm, {-\idx*\cellh-0.5*\cellh}) {\lbl};
}
\foreach \lbl/\idx in {Sonnet/0, Haiku/1, GPT-5/2, GPT-5m/3, Gem.~Pro/4, Gem.~Fl/5} {
  \node[font=\scriptsize\sffamily, rotate=45, anchor=south west, text=black!80]
    at ({\idx*\cellw+0.20cm}, 0.10cm) {\lbl};
}
\hcell{0}{0}{8.4}{s84}{white}
\hcell{1}{0}{5.6}{s55}{black!80}
\hcell{2}{0}{7.3}{s73}{white}
\hcell{3}{0}{6.1}{s60}{black!80}
\hcell{4}{0}{5.8}{s58}{black!80}
\hcell{5}{0}{5.8}{s58}{black!80}
\hcell{0}{1}{8.5}{s85}{white}
\hcell{1}{1}{6.4}{s64}{black!80}
\hcell{2}{1}{7.2}{s71}{white}
\hcell{3}{1}{6.9}{s69}{black!80}
\hcell{4}{1}{6.4}{s64}{black!80}
\hcell{5}{1}{6.0}{s60}{black!80}
\hcell{0}{2}{8.4}{s84}{white}
\hcell{1}{2}{7.3}{s73}{white}
\hcell{2}{2}{6.6}{s66}{black!80}
\hcell{3}{2}{6.8}{s68}{black!80}
\hcell{4}{2}{6.9}{s69}{black!80}
\hcell{5}{2}{6.5}{s64}{black!80}
\hcell{0}{3}{8.3}{s84}{white}
\hcell{1}{3}{6.4}{s64}{black!80}
\hcell{2}{3}{6.8}{s68}{black!80}
\hcell{3}{3}{6.8}{s68}{black!80}
\hcell{4}{3}{6.7}{s66}{black!80}
\hcell{5}{3}{6.8}{s68}{black!80}
\hcell{0}{4}{8.4}{s84}{white}
\hcell{1}{4}{7.1}{s71}{white}
\hcell{2}{4}{7.5}{s75}{white}
\hcell{3}{4}{6.8}{s68}{black!80}
\hcell{4}{4}{4.5}{s45}{black!80}
\hcell{5}{4}{6.7}{s66}{black!80}
\draw[black!30, rounded corners=3pt]
  (-0.02cm, 0.02cm) rectangle (6*\cellw+0.02cm, -5*\cellh-0.02cm);
\begin{scope}[yshift=-5*\cellh-0.65cm]
  \node[font=\tiny\sffamily, anchor=east, text=black] at (-0.05cm, 0) {4.5};
  \foreach \c/\idx in {s45/0, s55/1, s60/2, s64/3, s68/4, s73/5, s75/6, s84/7, s85/8} {
    \fill[\c, rounded corners=1pt] ({\idx*0.68cm}, -0.12cm) rectangle +({0.64cm}, 0.24cm);
  }
  \node[font=\tiny\sffamily, anchor=west, text=black] at ({9*0.68cm+0.05cm}, 0) {8.5};
  \node[font=\tiny\sffamily, text=black] at ({4.5*0.68cm}, -0.38cm) {Overall Score};
\end{scope}
\end{tikzpicture}
\caption{Format $\times$ model interaction heatmap (scores rounded to 1 decimal). Darker cells indicate higher scores. The two frontier models (Sonnet, GPT-5) show uniform color regardless of format; non-frontier models show wide variation --- the ``capability equalizer'' effect.}
\label{fig:heatmap}
\end{figure}
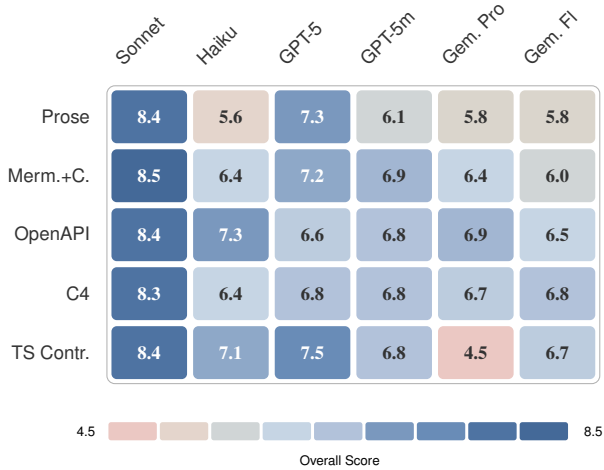

The two frontier models (Sonnet, GPT-5) show small format spreads ($0.17$ and $0.92$). All four non-frontier models show larger spreads, ranging from $0.83$ (GPT-5-mini) to $2.42$ (Gemini Pro). Gemini Pro's average score ($6.07$) places it closer to the mid-tier models than to the frontier pair, and its large spread is driven by a single anomalous condition (TS Contracts: $4.50$). OpenAI is the one family where frontier and mid-tier spreads are roughly equal ($0.92$ vs.\ $0.83$). Figure~\ref{fig:heatmap} visualizes this interaction.

\section{Agent Process Analysis}
\label{sec:process}

Beyond final code quality, our harness captures fine-grained process data, including every file write, compilation attempt, demo execution, and error encountered during each trial. This section analyzes \emph{how} agents build software under different specifications and model tiers, data rarely reported in code generation evaluations.

\subsection{Self-Validation Behavior}

The most deployment-relevant process metric is whether agents \emph{test their own output}. Our harness logs both TypeScript compilation attempts (\texttt{tsc --noEmit}) and end-to-end demo executions (\texttt{npx tsx src/demo.ts}). Table~\ref{tab:validation} presents validation behavior by model.

\begin{table}[htbp]
\centering
\caption{Agent Self-Validation Behavior by Model. TSC = TypeScript compiler invocations. Demo Run Rate = fraction of trials that executed the demo script at least once. All values are means across conditions.}
\label{tab:validation}
\begin{tblr}{
  colspec = {l c c c c},
  rowsep = 3pt,
  row{1} = {font=\bfseries, bg=tblHeaderTint},
  row{2} = {font=\bfseries, bg=tblHeaderTint},
  row{4,6,8} = {bg=tblRowAlt},
  hline{1,Z} = {0.8pt},
  hline{3} = {0.5pt},
  cell{3}{4} = {bg=demoHigh},
  cell{4}{4} = {bg=demoMedHigh},
  cell{5}{4} = {bg=demoMid},
  cell{6}{4} = {bg=demoMedLow},
  cell{7}{4} = {bg=demoLow},
  cell{8}{4} = {bg=demoNone},
}
{Model} & {TSC} & {TSC Pass} & {Demo Run} & {Demo} \\
 & {Attempts} & {Rate} & {Rate} & {Fail} \\
Sonnet 4.6 & 5.1 & 53\% & 100\% & 0.4 \\
Haiku 4.5 & 8.6 & 61\% & 80\% & 0.5 \\
GPT-5 & 4.1 & 13\% & 53\% & 0.7 \\
GPT-5-mini & 4.7 & 17\% & 40\% & 0.3 \\
Gemini 2.5 Pro & 3.7 & 22\% & 20\% & 0.3 \\
Gemini 2.5 Flash & 2.1 & 10\% & 0\% & 0.0 \\
\end{tblr}
\end{table}

Sonnet validated its output end-to-end in every trial. Gemini Flash \emph{never} ran the demo across any trial. It produced code and stopped without verification. Demo run rates decline monotonically across the capability spectrum (Table~\ref{tab:validation}, sorted by demo run rate). Agents that do not self-validate can ship code that compiles but fails at runtime.

\subsection{Compilation Effort}

Weaker models require more compilation iterations. Haiku attempted TSC $8.6$ times per trial versus Sonnet's $5.1$, but achieved a higher pass rate ($61\%$ vs.\ $53\%$), indicating Haiku makes more incremental fixes per attempt. GPT-5-mini had the worst compilation efficiency at $4.7$ attempts with only a $17\%$ pass rate, often introducing new errors while fixing old ones.

Specification format moderates compilation effort. On Haiku, structured specifications reduced TSC failures from $3.33$ (prose mean) to $2.33$ (Mermaid+constraints). Explicit component boundaries help the agent produce correct code on the first attempt.

\subsection{Debugging Intensity}

We define \emph{rewrite rate} as the fraction of file writes that overwrite a previously written file, a proxy for debugging intensity. Table~\ref{tab:rewrite} shows rewrite rates by model and format.

\begin{table}[htbp]
\centering
\caption{File Rewrite Rate (\% of writes that are overwrites). A dash indicates fewer than 3 file writes in that condition.}
\label{tab:rewrite}
\begin{tblr}{
  colspec = {l c c c c},
  rowsep = 3pt,
  row{1} = {font=\bfseries, bg=tblHeaderTint},
  row{3,5} = {bg=tblRowAlt},
  hline{1,Z} = {0.8pt},
  hline{2} = {0.5pt},
}
{Format} & {Sonnet} & {Haiku} & {GPT-mini} & {Gem.\ Fl.} \\
Prose & 15\% & 20\% & 11\% & 33\% \\
Mermaid+Constr. & 36\% & 9\% & 17\% & 20\% \\
OpenAPI & 15\% & 10\% & 8\% & 13\% \\
C4 & 28\% & 16\% & --- & 23\% \\
TS Contracts & 23\% & 17\% & --- & --- \\
\end{tblr}
\end{table}

Two patterns emerge. First, Gemini Flash with prose has the highest rewrite rate ($33\%$). One in three file writes is a correction, indicating the agent is floundering. OpenAPI reduces this to $13\%$, as the formal API contract gives the agent a clearer implementation target. Second, Sonnet with Mermaid+constraints has a high rewrite rate ($36\%$) despite high final quality ($8.50$). This corresponds to the trials where Sonnet spent $40+$ turns iterating toward closer specification adherence, rewriting files to better match the specification.

\subsection{Code Volume and Completeness}

Models differ in code output volume:

\begin{itemize}
    \item Sonnet 4.6: $1{,}504$ lines mean, $14.3$ files
    \item Haiku 4.5: $1{,}350$ lines, $15.7$ files
    \item GPT-5-mini: $642$ lines, $12.1$ files
    \item Gemini Flash: $590$ lines, $9.7$ files
\end{itemize}

GPT-5-mini and Gemini Flash produce less than half the code of Sonnet, yet their completeness scores are only $1$--$2$ points lower ($7.0$ vs.\ $9.0$). The judge evaluates architectural correctness rather than implementation depth. A smaller but well-structured implementation scores better than a verbose but architecturally flawed one.

\subsection{Automated Architecture Verification}

To complement the subjective judge scores, we validated each trial against the architecture specification using two automated metrics: API route coverage (percentage of 25 specified routes with a handler in the generated router) and a weighted architecture compliance score combining component existence (20\%), communication patterns (20\%), behavioral constraints (20\%), route coverage (20\%), data ownership (10\%), and file structure (10\%). Approximately 80\% of the architecture is machine-verifiable. Unchecked elements include design rationale, full status transition matrix, and semantic data ownership. Table~\ref{tab:arch_verification} presents both metrics.

\begin{table*}[htbp]
\centering
\caption{Automated Architecture Verification by Model and Format. All values are percentages. \textsc{Route} = API route coverage (25 routes). \textsc{Compl} = weighted compliance (Components 20\%, Communication 20\%, Constraints 20\%, Routes 20\%, Data 10\%, Structure 10\%). Covers $\sim$80\% of the architecture.}
\label{tab:arch_verification}
\begin{tblr}{
  colspec = {X c c c c c c c c c c c c},
  width = \textwidth,
  rowsep = 3pt,
  row{1} = {font=\bfseries, bg=tblHeaderTint},
  row{2} = {font=\bfseries, bg=tblHeaderTint},
  row{3} = {font=\bfseries, bg=tblHeaderTint},
  hline{1,Z} = {0.8pt},
  hline{4} = {0.5pt},
  vline{4,6,8,10,12} = {2-Z}{0.3pt, gray},
  cell{4}{2} = {bg=bestcell}, cell{4}{4} = {bg=bestcell},
  cell{4}{6} = {bg=bestcell}, cell{4}{8} = {bg=bestcell},
  cell{4}{10} = {bg=bestcell}, cell{4}{12} = {bg=worstcell},
  cell{5}{2} = {bg=bestcell}, cell{5}{4} = {bg=bestcell},
  cell{5}{6} = {bg=bestcell}, cell{5}{8} = {bg=bestcell},
  cell{5}{10} = {bg=bestcell}, cell{5}{12} = {bg=worstcell},
  cell{6}{2} = {bg=bestcell}, cell{6}{4} = {bg=bestcell},
  cell{6}{6} = {bg=bestcell}, cell{6}{8} = {bg=bestcell},
  cell{6}{10} = {bg=bestcell}, cell{6}{12} = {bg=worstcell},
  cell{7}{2} = {bg=bestcell}, cell{7}{4} = {bg=bestcell},
  cell{7}{6} = {bg=bestcell}, cell{7}{8} = {bg=bestcell},
  cell{7}{10} = {bg=tblAmber}, cell{7}{12} = {bg=tblAmber},
  cell{8}{2} = {bg=bestcell}, cell{8}{4} = {bg=bestcell},
  cell{8}{6} = {bg=bestcell}, cell{8}{8} = {bg=bestcell},
  cell{8}{10} = {bg=bestcell}, cell{8}{12} = {bg=bestcell},
}
 & \SetCell[c=4]{c}{Anthropic} & & & & \SetCell[c=4]{c}{OpenAI} & & & & \SetCell[c=4]{c}{Google} & & & \\
 & \SetCell[c=2]{c}{Sonnet} & & \SetCell[c=2]{c}{Haiku} & & \SetCell[c=2]{c}{GPT-5} & & \SetCell[c=2]{c}{GPT-mini} & & \SetCell[c=2]{c}{Gem.\ Pro} & & \SetCell[c=2]{c}{Gem.\ Flash} & \\
{Format} & \textsc{Route} & \textsc{Compl} & \textsc{Route} & \textsc{Compl} & \textsc{Route} & \textsc{Compl} & \textsc{Route} & \textsc{Compl} & \textsc{Route} & \textsc{Compl} & \textsc{Route} & \textsc{Compl} \\
Prose & 100 & 100 & 100 & 88 & 100 & 100 & 85 & 97 & 100 & 93 & 33 & 71 \\
Mermaid+C. & 100 & 97 & 100 & 93 & 100 & 100 & 100 & 98 & 100 & 95 & 33 & 73 \\
OpenAPI & 100 & 100 & 100 & 97 & 100 & 100 & 100 & 100 & 100 & 100 & 33 & 72 \\
C4 & 100 & 97 & 100 & 97 & 100 & 100 & 100 & 98 & 67 & 85 & 67 & 85 \\
TS Contr. & 100 & 100 & 100 & 100 & 100 & 100 & 100 & 100 & 100 & 100 & \textbf{100} & \textbf{100} \\
\end{tblr}
\end{table*}

Route coverage reveals a pattern invisible in judge scores. Gemini Flash implements only one-third of specified routes under prose, Mermaid, and OpenAPI, but achieves 100\% coverage with TypeScript contracts, a $3\times$ improvement from format alone. The typed interface signatures gave the weakest model an unambiguous implementation checklist that diagrams and formal API schemas did not. Sonnet, Haiku, and GPT-5 achieve 100\% regardless of format. GPT-5-mini drops to 85\% on prose but maintains 100\% with any structured format.

The compliance score confirms the same pattern at a higher level. TypeScript contracts is the only format that achieves $100\%$ compliance across all six models from all three vendor families. Every other format has at least one model where compliance falls below $97\%$. The weakest model in the experiment, which never runs its own demo, produces half the code, and stops after 12 turns, achieves perfect automated architecture compliance when given typed interface contracts. The same model, given the same architectural information as prose, misses a third of the API routes and scores below $71\%$.

Mermaid diagrams require interpreting visual relationships. C4 models require mapping hierarchical decomposition to code. OpenAPI specifies routes but not internal structure. TypeScript interfaces are \emph{already code}. The model can implement the interfaces directly without translating between representations. BaxBench~\cite{vero2025baxbench} showed that OpenAPI outperforms prose for API-level correctness. TypeScript contracts go further by specifying both the API surface and the internal service boundaries. For a model with limited architectural reasoning capacity, eliminating this translation step is the difference between a complete and an incomplete implementation.

GPT-5 achieves $100\%$ on every format and does not need structured specifications at all. The frontier OpenAI model implicitly extracts the same architectural structure from prose that the weakest Google model can only extract from typed interfaces. The structured specification compensates for exactly the capability gap between these two extremes.

\subsection{Resource Efficiency}

Table~\ref{tab:efficiency} presents the cost--quality tradeoff across models, using total tokens as a proxy for API cost.

\begin{table}[htbp]
\centering
\caption{Resource Efficiency by Model}
\label{tab:efficiency}
\begin{tblr}{
  colspec = {l c c c},
  rowsep = 3pt,
  row{1} = {font=\bfseries, bg=tblHeaderTint},
  row{3,5,7} = {bg=tblRowAlt},
  hline{1,Z} = {0.8pt},
  hline{2} = {0.5pt},
  cell{6}{4} = {bg=worstcell},
  cell{7}{4} = {bg=bestcell},
}
{Model} & {Tokens (K)} & {Overall} & {Score/100K} \\
Sonnet 4.6 & 640 & 8.42 & 1.32 \\
GPT-5 & 248 & 7.08 & 2.85 \\
Gemini Pro & 329 & 6.07 & 1.85 \\
Haiku 4.5 & 735 & 6.50 & 0.88 \\
GPT-5-mini & 225 & 6.72 & 2.99 \\
Gemini Flash & 223 & 6.35 & 2.85 \\
\end{tblr}
\end{table}

GPT-5-mini and Gemini Flash achieve the highest score-per-token ratios despite lower absolute quality, because they consume $3$--$4\times$ fewer tokens. Haiku is the least efficient. It consumes \emph{more} tokens than Sonnet ($735$K vs.\ $640$K) while scoring $1.9$ points lower, because it spends tokens on extended compilation debugging loops that Sonnet avoids. The ``cheaper'' model is not always cheaper in practice when measured by total tokens to completion.

\subsection{Failure Mode Taxonomy}

The process data reveals three distinct failure modes.

\begin{enumerate}
    \item Compilation death spiral (Haiku, GPT-5-mini). The agent enters a fix loop where correcting one TypeScript error introduces another, consuming turns without converging. Haiku averaged $8.6$ TSC attempts per trial.
    \item Premature termination (Gemini Flash). The agent writes files and stops after $12.5$ turns mean without compiling or testing. It produces structurally complete but unverified code.
    \item Perfectionist iteration (Sonnet with structured specs). The agent achieves compilable code early but continues rewriting to better match the specification, consuming $38$+ turns and $1{,}000$K+ tokens. This produces the highest quality but at disproportionate cost.
\end{enumerate}

These failure modes interact with specification format. Structured specifications can trigger perfectionist iteration on strong models (increasing cost without proportional quality gain) while reducing compilation death spirals on weak models by providing clearer implementation targets.

\section{Discussion}
\label{sec:discussion}

\subsection{Does Architecture Guidance Matter at All?}

Before comparing specification formats, we validate the premise that architecture guidance is beneficial. We ran three additional trials with Sonnet 4.6 using a requirements-only prompt: identical features, API routes, and data fields, but zero architectural guidance --- no components, no event bus, no service boundaries, no data ownership rules, no file structure. Table~\ref{tab:no_arch} compares the results.

\begin{table}[htbp]
\centering
\caption{Architecture Guidance vs.\ Requirements Only (Sonnet 4.6, 3 trials each). No Arch = functional requirements only; Prose = lightest architecture specification. Both use identical tooling.}
\label{tab:no_arch}
\begin{tblr}{
  colspec = {l c c c},
  rowsep = 3pt,
  row{1} = {font=\bfseries, bg=tblHeaderTint},
  row{4,6} = {bg=tblRowAlt},
  row{10,11} = {bg=tblRowAlt},
  row{14} = {bg=tblRowAlt},
  hline{1,Z} = {0.8pt},
  hline{2} = {0.5pt},
  cell{2}{1} = {c=4}{l,font=\itshape,bg=groupheader},
  cell{8}{1} = {c=4}{l,font=\itshape,bg=groupheader},
  cell{12}{1} = {c=4}{l,font=\itshape,bg=groupheader},
  hline{2,8,12} = {0.5pt},
  hline{3,9,13} = {0.3pt},
}
{Metric} & {No Arch} & {Prose} & {$\Delta$} \\
{Judge Scores (1--10)} & & & \\
Arch.\ Adherence & $3.00$ & $7.67$ & $-4.67$ \\
Completeness & $7.00$ & $9.00$ & $-2.00$ \\
Code Quality & $7.33$ & $9.00$ & $-1.67$ \\
Constraint Compliance & $3.00$ & $8.00$ & $-5.00$ \\
Overall & $5.08$ & $8.42$ & $-3.34$ \\
{Resource Consumption} & & & \\
Tokens (K) & $436$ & $433$ & $+3$ \\
Components (/9) & $8.0$ & $9.0$ & $-1$ \\
Violations & $0$ & $0$ & $0$ \\
{Automated Architecture Verification} & & & \\
Route Coverage & $100\%$ & $100\%$ & $0$ \\
Compliance Score & $96.8\%$ & $100\%$ & $-3.2$ \\
\end{tblr}
\end{table}

The gap is large and consistent. Architecture guidance raises the overall judge score by $3.34$ points ($5.08 \to 8.42$), with the steepest drops in architectural adherence ($-4.67$) and constraint compliance ($-5.00$). Without being told about service isolation or event-driven communication, the agent builds a functionally complete system (100\% route coverage, working demo) but defaults to a flat handler-and-shared-store pattern. None of the three trials produced an event bus. The agent satisfies every functional requirement but invents no architecture.

This confirms the premise stated in the Introduction: the format comparison that follows rests on a foundation where architecture guidance itself accounts for a $3+$-point quality improvement. The remaining sections examine how to deliver that guidance most effectively.

\subsection{Structured Specs and the Capability Gap}

The 90-trial cross-vendor comparison confirms the paper's central finding. Structured specifications disproportionately benefit weaker models. The two frontier models (Sonnet, GPT-5) show format spreads of $0.17$ and $0.92$. The four non-frontier models show larger spreads of $0.83$--$2.42$. The pattern holds cleanly for Anthropic (Sonnet $0.17$ vs.\ Haiku $1.67$, a $10\times$ difference). OpenAI shows roughly equal spreads across tiers ($0.92$ vs.\ $0.83$), suggesting that format sensitivity may plateau for models above a certain capability threshold.

\begin{figure}[htbp]
\centering
\begin{tikzpicture}
\begin{axis}[
  ybar,
  width=\columnwidth,
  height=5.2cm,
  bar width=13pt,
  ymin=5, ymax=9.5,
  ytick={5,6,7,8,9},
  ylabel={Overall Score},
  ylabel style={font=\small\sffamily},
  yticklabel style={font=\scriptsize\sffamily},
  xtick=data,
  symbolic x coords={Sonnet,Haiku,GPT-5m,Gem.\ Pro,Gem.\ Fl},
  xticklabel style={font=\scriptsize\sffamily, rotate=30, anchor=east},
  legend style={font=\scriptsize\sffamily, at={(0.98,0.97)}, anchor=north east,
    legend columns=1, draw=gray!30, fill=white,
    /tikz/every even column/.append style={column sep=3pt}},
  area legend,
  enlarge x limits=0.12,
  ymajorgrids=true,
  grid style={gray!15, dashed},
  axis lines*=left,
  axis line style={gray!60},
  tick style={gray!60},
]
\addplot[fill=figEvalFill, draw=figInputBorder, line width=0.5pt]
  coordinates {
    (Sonnet,8.42)
    (Haiku,5.58)
    (GPT-5m,6.08)
    (Gem.\ Pro,5.83)
    (Gem.\ Fl,5.75)
  };
\addplot[fill=figBarFill, draw=figAgentBorder, line width=0.5pt]
  coordinates {
    (Sonnet,8.50)
    (Haiku,7.25)
    (GPT-5m,6.92)
    (Gem.\ Pro,6.92)
    (Gem.\ Fl,6.83)
  };
\legend{Prose, Best format}
\end{axis}
\end{tikzpicture}
\caption{The capability equalizer effect. Under prose (light bars), quality drops steeply from frontier to weaker models. The best structured format per model (dark bars) lifts non-frontier scores, narrowing the gap from $2.8$ to $1.7$ points.}
\label{fig:equalizer}
\end{figure}
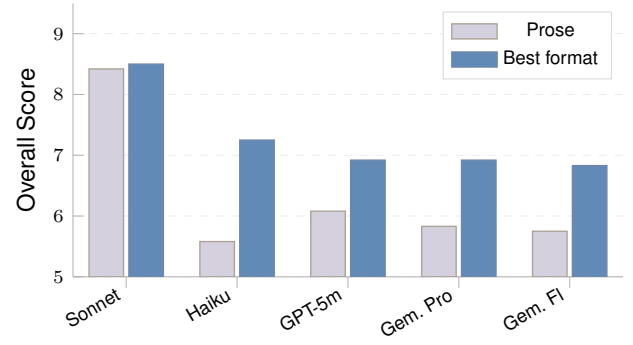

Figure~\ref{fig:equalizer} visualizes this effect. Under prose, quality drops steeply from frontier to weaker models. Under structured formats, the weaker models are lifted toward the frontier --- the specification narrows the capability gap.

The effect is strongest when measured by route coverage rather than judge scores. Gemini Flash implements only $33\%$ of specified API routes under prose but $100\%$ under TypeScript contracts. This $3\times$ improvement from format alone, on the same model with the same information, is the experiment's most practically significant finding.

\subsection{No Universal Best Format}

The best format varies by model. Mermaid+constraints leads on Sonnet ($8.50$), OpenAPI on Haiku ($7.25$) and Gemini Pro ($6.92$), TypeScript contracts on GPT-5 ($7.50$), and C4 on Gemini Flash ($6.83$). No single structured format dominates across all models.

Format effectiveness depends on model-specific strengths, consistent with ConCodeEval's finding that schema format significantly affects constraint satisfaction even at the data level~\cite{kammakomati2024concodeeval}. Claude models respond well to visual component diagrams (Mermaid). GPT-5 responds to typed interface contracts. Gemini models respond to hierarchical decomposition (C4) and formal API schemas (OpenAPI). Teams should test their specific model against multiple formats rather than assuming one structured format is universally optimal.

\subsection{Inverted Cost Efficiency}

Haiku 4.5 consumes $735$K tokens per trial, $15\%$ more than Sonnet's $640$K, while scoring $1.9$ points lower. Haiku spends tokens on repeated compilation attempts ($8.6$ per trial vs.\ Sonnet's $5.1$) and extended debugging loops. The ``cheaper'' model is not cheaper when measured by total tokens to completion.

This inversion does not hold for GPT-5-mini ($225$K, score $6.72$) or Gemini Flash ($223$K, score $6.35$), which are both fast and cheap but achieve lower quality through premature termination rather than extended debugging. The cost inversion is specific to models capable enough to attempt iterative repair but not capable enough to succeed quickly, a ``valley'' in the capability--cost curve. Dente et al.~\cite{dente2026constraint} observed a related phenomenon: adding constraints degraded performance most for mid-capability models that attempted but failed to satisfy them.

\subsection{Self-Validation as a Capability Indicator}

Demo run rate (whether the agent tests its own output end-to-end) is an effective proxy for model capability. Sonnet runs demos in $100\%$ of trials. The rate declines monotonically: Haiku $80\%$, GPT-5 $53\%$, GPT-5-mini $40\%$, Gemini Pro $20\%$, Gemini Flash $0\%$. This metric requires no judge, no subjective scoring, and no ground truth. It is a fully objective behavioral signal that correlates strongly with final quality.

A practical guardrail follows. If an agent does not run its own tests, the generated code should not be trusted without external validation.

\subsection{Can We Trust the Judge?}

To validate our LLM-as-judge evaluation, we conducted two analyses: correlation with automated compliance metrics, and cross-vendor inter-judge agreement using a second judge model (GPT-5) from a different vendor family.

\subsubsection{Judge--Automation Correlation}

Table~\ref{tab:judge_corr} compares how well each judge's scores align with automated metrics (Sonnet across all 90 trials; GPT-5 across the 28 of 30 re-judged trials that returned parseable scores).

\begin{table}[htbp]
\centering
\caption{Judge--Automation Correlation (Pearson $r$)}
\label{tab:judge_corr}
\begin{tblr}{
  colspec = {l l c c},
  rowsep = 3pt,
  row{1} = {font=\bfseries, bg=tblHeaderTint},
  row{3,5} = {bg=tblRowAlt},
  hline{1,Z} = {0.8pt},
  hline{2} = {0.5pt},
}
{Dimension} & {Auto Metric} & {Sonnet ($n$=90)} & {GPT-5 ($n$=28)} \\
Completeness & Route coverage & $0.63$ & $0.85$ \\
Constraint Compl. & Violations & $-0.33$ & $-0.48$ \\
Constraint Compl. & Compliance \% & $0.21$ & $-0.02$ \\
Arch.\ Adherence & Compliance \% & $0.15$ & $0.15$ \\
\end{tblr}
\end{table}

GPT-5 correlates more strongly with automated metrics than Sonnet on two key dimensions: completeness ($r = 0.85$ vs.\ $0.63$) and constraint violations ($r = -0.48$ vs.\ $-0.33$). Both judges correlate weakly with the weighted compliance score for architecture, indicating that all judges evaluate architectural quality on a different axis than import-level static analysis.

In 11 of 90 Sonnet-judged trials, the automated checker found zero violations but the judge scored constraint compliance $\leq 5$. The judge caught semantic violations invisible to import analysis: router-mediated cross-service orchestration, constructor-injected dependencies that bypass the event bus. In one trial, the judge gave a perfect $10$ despite a detected cross-service import.

\subsubsection{Inter-Judge Agreement}

We re-judged 30 trials (5 per model, one per specification format) with GPT-5 to test whether scores depend on the judge model. Table~\ref{tab:interjudge} presents the results.

\begin{table}[htbp]
\centering
\caption{Inter-Judge Agreement: Sonnet 4.6 vs GPT-5 ($n$=30). Bias = GPT-5 mean $-$ Sonnet mean. Positive = GPT-5 more lenient.}
\label{tab:interjudge}
\begin{tblr}{
  colspec = {l c c c c},
  rowsep = 3pt,
  row{1} = {font=\bfseries, bg=tblHeaderTint},
  row{3,5} = {bg=tblRowAlt},
  row{7} = {font=\bfseries, bg=tblAvgRow},
  hline{1,Z} = {0.8pt},
  hline{2} = {0.5pt},
  hline{6} = {0.3pt},
}
{Dimension} & {Sonnet mean} & {GPT-5 mean} & {Bias} & {$r$} \\
Arch.\ Adherence & $6.53$ & $7.03$ & $+0.50$ & $0.51$ \\
Completeness & $7.37$ & $7.70$ & $+0.33$ & $0.60$ \\
Code Quality & $7.60$ & $7.17$ & $-0.43$ & $0.52$ \\
Constraint Compl. & $6.50$ & $7.77$ & $+1.27$ & $0.37$ \\
Overall & $7.00$ & $7.42$ & $+0.42$ & $0.47$ \\
\end{tblr}
\end{table}

The two judges agree moderately on relative rankings ($r = 0.47$--$0.60$) but differ in absolute calibration. GPT-5 is uniformly more lenient, scoring $+0.42$ points higher on average. The largest disagreement is on constraint compliance ($+1.27$), where GPT-5 grades more generously than Sonnet.

\subsubsection{Same-Family Bias}

A known concern with LLM-as-judge is self-preference bias, where models score their own family's code higher. We tested this by comparing the GPT-5$-$Sonnet bias across code generated by each vendor family (Table~\ref{tab:bias}).

\begin{table}[htbp]
\centering
\caption{Same-Family Bias Test: GPT-5 $-$ Sonnet Overall Score}
\label{tab:bias}
\begin{tblr}{
  colspec = {l c c c},
  rowsep = 3pt,
  row{1} = {font=\bfseries, bg=tblHeaderTint},
  row{3} = {bg=tblRowAlt},
  hline{1,Z} = {0.8pt},
  hline{2} = {0.5pt},
}
{Generator Family} & {Sonnet judge} & {GPT-5 judge} & {Bias} \\
Anthropic ($n$=10) & $7.55$ & $8.22$ & $+0.67$ \\
OpenAI ($n$=8) & $6.94$ & $7.94$ & $+1.00$ \\
Google ($n$=10) & $6.60$ & $7.67$ & $+1.08$ \\
\end{tblr}
\end{table}

If same-family bias existed, Sonnet would score Anthropic-generated code higher than GPT-5 does (negative bias for Anthropic rows), and GPT-5 would score OpenAI-generated code higher (larger positive bias for OpenAI rows). Neither pattern holds. GPT-5 is uniformly more lenient across all three families, with the smallest bias on Anthropic code ($+0.67$) and the largest on Google code ($+1.08$). The observed pattern is leniency bias, not same-family preference.

This cross-judge validation strengthens the paper's findings in two ways. First, the relative rankings produced by both judges are consistent ($r = 0.47$--$0.60$), so the format $\times$ model interaction holds regardless of which judge is used. Second, the absence of same-family bias indicates that our primary results (judged by Sonnet) are not inflated for Sonnet-generated code relative to other vendors.

\subsection{Process Metrics as First-Class Evaluation}

Code generation benchmarks overwhelmingly report outcome metrics (pass@k, judge scores)~\cite{jimenez2024swebench, chen2021codex}. \emph{How} agents build software is at least as informative as the final quality score. Two models can achieve the same score through radically different processes. Sonnet writes files methodically and validates iteratively. Gemini Flash writes files quickly and stops. The process data (TSC pass rate, demo run rate, rewrite rate, turn count) provides deployment-relevant information that outcome scores cannot.

\subsection{Specification as Implementation Interface}

The TypeScript contracts result --- the only format achieving $100\%$ route coverage across all six models --- suggests that the most effective architecture specification is one that is already code. TypeScript interfaces do not require the agent to translate between a human-readable description and an implementation structure; the specification \emph{is} the structure. This parallels Wang et al.'s CodeSpec~\cite{wang2026codespec}, which found that executable specifications (LLM-generated checkers) outperform textual ones by $28$ percentage points. Our finding extends this from generated checkers to human-authored interface contracts. The implication for practitioners is that architecture specifications for coding agents should be \emph{compilable}, not merely parseable. As the boundary between specification and implementation collapses, the traditional distinction between ``design document'' and ``code skeleton'' may become a liability rather than an abstraction benefit.

\subsection{Practical Recommendations}

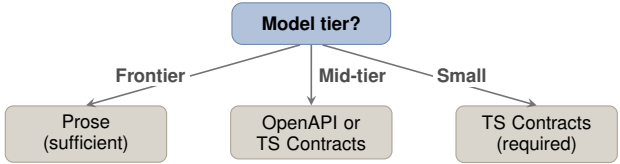
\begin{figure}[htbp]
\centering
\begin{tikzpicture}[
  every node/.style={font=\scriptsize\sffamily},
  qbox/.style={draw=figAgentBorder, rounded corners=4pt, fill=figAgentFill,
    minimum width=2.1cm, minimum height=0.55cm, align=center,
    font=\scriptsize\sffamily\bfseries},
  abox/.style={draw=figInputBorder, rounded corners=3pt, fill=figInputFill,
    minimum width=2.0cm, minimum height=0.5cm, align=center,
    text width=1.9cm, font=\scriptsize\sffamily},
  rbox/.style={draw=figEvalBorder, rounded corners=3pt, fill=figEvalFill,
    minimum width=2.0cm, minimum height=0.5cm, align=center,
    text width=1.9cm, font=\scriptsize\sffamily},
  arr/.style={->, >=stealth, line width=0.6pt, black!55},
  elbl/.style={font=\scriptsize\sffamily\bfseries, text=black!70, fill=white, inner sep=1.5pt},
]
\node[qbox] (q1) {Model tier?};
\node[abox, below left=0.8cm and 0.85cm of q1] (a1) {Prose\\(sufficient)};
\node[abox, below=0.8cm of q1] (a2) {OpenAPI or\\TS Contracts};
\node[abox, below right=0.8cm and 0.85cm of q1] (a3) {TS Contracts\\(required)};
\draw[arr] (q1.south) ++(-0.15,0) -- (a1.north) node[elbl, midway, left=1pt] {Frontier};
\draw[arr] (q1.south) -- (a2.north) node[elbl, midway, right=1pt] {Mid-tier};
\draw[arr] (q1.south) ++(0.15,0) -- (a3.north) node[elbl, midway, right=1pt] {Small};
\end{tikzpicture}
\caption{Specification format decision guide. Frontier models (Sonnet, GPT-5) need no structured specification. Mid-tier models benefit from OpenAPI or TypeScript contracts. Small models require TypeScript contracts for complete route coverage.}
\label{fig:decision}
\end{figure}

The optimal specification strategy depends on model tier (Figure~\ref{fig:decision}).

\begin{itemize}
    \item Frontier models (Sonnet, GPT-5). Prose is sufficient and most cost-efficient. Structured specs yield marginal quality gains at significant token overhead.
    \item Mid-tier models (Haiku, GPT-5-mini, Gemini Pro). Structured specifications provide measurable improvement ($0.83$--$2.42$ point spreads). OpenAPI or TypeScript contracts are the safest choices across vendors.
    \item Small models (Gemini Flash). Structured specifications are essential. TypeScript interface contracts are the only format that achieves $100\%$ route coverage.
    \item All models. Include a numbered constraint list. Constraint compliance is the dimension most sensitive to both format and model capability. Monitor demo run rate as a deployment readiness signal.
\end{itemize}

\subsection{Limitations}

Sample size and statistical power. With $n=3$ per cell across a $5 \times 6$ design, individual cell comparisons are underpowered and we do not report statistical significance tests. We frame the format $\times$ model interaction as an exploratory finding supported by consistent directional patterns across vendors. The strongest evidence comes from automated metrics (route coverage, compliance scores) that are deterministic census measurements rather than samples --- the $33\% \rightarrow 100\%$ route coverage improvement requires no statistical test. Larger samples ($n \geq 5$) per cell would enable formal hypothesis testing.

No open-source models. All six models are proprietary (Anthropic, OpenAI, Google). Open-source models (Llama, Mistral) may exhibit different interaction patterns, particularly at the lower capability tier.

Single system. We tested one system of moderate complexity. Larger systems may amplify format effects even on frontier models.

Judge consistency. Using Sonnet 4.6 as judge for all rounds ensures comparability but introduces potential same-family bias for Sonnet-generated code. Self-preference bias is documented for GPT-4~\cite{zheng2023judging}. Its magnitude for Claude is unknown.

Format coverage. Our five formats were chosen for proximity to software engineering practice with coding agents. We did not test systems engineering formats such as UML, SysML v2, AADL (Architecture Analysis \& Design Language), or cloud infrastructure languages such as AWS CloudFormation or Terraform, which describe architecture at different abstraction levels. Whether these formats exhibit the same capability-equalizer effect is an open question, particularly for models trained on corpora where these formats are less prevalent than TypeScript or OpenAPI.

Specification quality. All specifications were written by the same authors. The prose may be unusually clear or the structured formats unusually well-organized.

\section{Threats to Validity}
\label{sec:threats}

Internal validity. All specifications were written by the same authors, introducing potential bias. We mitigated this by reviewing all five for information equivalence. The six model rounds were run sequentially rather than interleaved, so temporal effects (API load, model updates) could confound the comparison. Each trial starts from a clean context with no shared state.

External validity. We tested one system, six models from three families, and five formats. The system's moderate complexity ($\sim$7 components) may understate advantages that emerge at larger scales. Generalization to other domains (data pipelines, UI applications) requires further study.

Construct validity. LLM-as-judge evaluation does not perfectly correlate with human expert assessment, though recent work shows strong correlation~\cite{zheng2023judging}. Using Sonnet as judge for Sonnet-generated code could introduce same-family bias absent when judging Haiku-generated code.

\section{Future Work}
\label{sec:future}

\begin{itemize}
    \item Model capability continuum. Testing additional model tiers (e.g., GPT-4o vs.\ GPT-4o-mini, Gemini Pro vs.\ Flash) to map the capability threshold at which format effects emerge.
    \item Hybrid specifications. Testing combinations of formats (e.g., OpenAPI + TypeScript interfaces + constraint list) based on the dimension-specific and model-specific strengths reported above.
    \item Scaling. Evaluating the effect at larger system scales (10+ services). We hypothesize the format $\times$ model interaction strengthens with system complexity.
    \item Variance reduction. Investigating why structured formats produce higher variance and whether techniques like self-verification reduce it.
    \item Automated constraint enforcement. Integrating architectural constraint checking into the agent's tool loop, testing whether real-time feedback reduces the model capability dependency.
    \item Incremental development. Measuring whether a second agent can extend the first agent's code while preserving architectural constraints under each format.
\end{itemize}

\section{Conclusion}
\label{sec:conclusion}

We compared five architecture specification formats across six models from three vendor families in 90 multi-turn agent trials ($5 \times 6$ factorial design). The central finding is a strong \emph{format $\times$ model interaction}. On the two frontier models (Sonnet, GPT-5), specification format barely matters (spread $0.17$--$0.92$). On the four non-frontier models, format produces quality spreads of $0.83$--$2.42$ points. Code-proximate formats (OpenAPI, TypeScript contracts) are the most robust to model degradation, and TypeScript contracts triple API route coverage for the weakest model ($33\% \rightarrow 100\%$).

Process analysis reveals that mid-tier models can consume more tokens than frontier models for worse output when trapped in compilation debugging loops, that self-validation rates collapse from $100\%$ to $0\%$ across the capability spectrum, and that distinct failure modes (compilation death spirals, premature termination) interact differently with specification format. These process-level findings are rarely reported in code generation evaluations but are critical for deployment decisions.

Structured architecture specifications do not primarily improve the best achievable quality. Frontier models handle prose well. They function as a \emph{capability equalizer}, with value inversely proportional to model strength. As Konrad et al.~\cite{konrad2026architecture} argue, AI coding agents need architectural governance. The specification format is itself a governance mechanism, and its effectiveness depends on the capability of the agent being governed. This parallels a familiar pattern in human organizations: junior developers need more detailed specifications than senior ones. Organizations deploying weaker or cost-optimized models must invest proportionally more in specification quality --- the specification budget is not optional overhead but a direct lever on output quality. As LLM coding agents are deployed across a widening range of model capabilities and cost points, architecture specification format becomes a first-order engineering decision with measurable quality and cost consequences.

\section*{Data Availability}

All 93 trial transcripts, generated codebases, judge scores, automated compliance results, specification files, and the experiment harness are publicly available at \url{https://github.com/arquicanedo/architecture-as-equalizer}.


\bibliographystyle{IEEEtran}

\begin{thebibliography}{99}

\bibitem{vero2025baxbench}
M.~Vero, N.~Mundler, V.~Chibotaru, V.~Raychev, M.~Baader, N.~Jovanovic, J.~He, and M.~Vechev,
``BaxBench: Can LLMs generate correct and secure backends?''
\emph{arXiv preprint arXiv:2502.11844}, 2025.

\bibitem{dente2026constraint}
F.~Dente, D.~Satriani, and P.~Papotti,
``Constraint Decay: The fragility of LLM agents in backend code generation,''
\emph{arXiv preprint arXiv:2605.06445}, 2026.

\bibitem{wang2026codespec}
P.~Wang, L.~Zhang, F.~Liu, T.~Li, and Y.~Zhu,
``CodeSpec: Dual executable specifications for agentic long-horizon feature development,''
\emph{arXiv preprint arXiv:2607.26777}, 2026.

\bibitem{meawad2026designfirst}
F.~Meawad,
``Design-first governance for AI-generated code,''
in \emph{Proc.\ IEEE ICSA Companion}, 2026.
\textit{(Abstract only; full text behind IEEE paywall.)}

\bibitem{chen2021codex}
M.~Chen et al.,
``Evaluating large language models trained on code,''
\emph{arXiv preprint arXiv:2107.03374}, 2021.

\bibitem{jimenez2024swebench}
C.~E.~Jimenez, J.~Yang, A.~Wettig, S.~Yao, K.~Pei, O.~Press, and K.~Narasimhan,
``SWE-bench: Can language models resolve real-world GitHub issues?''
in \emph{Proc.\ ICLR}, 2024.

\bibitem{wei2022chain}
J.~Wei et al.,
``Chain-of-thought prompting elicits reasoning in large language models,''
in \emph{Proc.\ NeurIPS}, 2022.

\bibitem{jiang2024selfplanning}
X.~Jiang et al.,
``Self-planning code generation with large language models,''
\emph{ACM Trans.\ Softw.\ Eng.\ Methodol.}, 2024.

\bibitem{perry1992foundations}
D.~E.~Perry and A.~L.~Wolf,
``Foundations for the study of software architecture,''
\emph{ACM SIGSOFT Softw.\ Eng.\ Notes}, vol.~17, no.~4, pp.~40--52, 1992.

\bibitem{terra2012qualitas}
R.~Terra, M.~T.~Valente, K.~Czarnecki, and R.~S.~Bigonha,
``Recommending refactorings to reverse software architecture erosion,''
in \emph{Proc.\ CSMR}, 2012, pp.~335--340.

\bibitem{archunit}
P.~Gafert,
``ArchUnit: Unit test your Java architecture,''
\url{https://www.archunit.org/}, 2017.

\bibitem{bogner2024llmarchitecture}
J.~Bogner, M.~Merkel, and S.~Wagner,
``On the architecture-level quality of LLM-generated code,''
in \emph{Proc.\ IEEE ICSA}, 2024.

\bibitem{mermaid}
K.~Sveidqvist,
``Mermaid: Generation of diagrams and flowcharts from text,''
\url{https://mermaid.js.org/}, 2014.

\bibitem{plantuml}
A.~Roques,
``PlantUML,''
\url{https://plantuml.com/}, 2009.

\bibitem{brown2018c4}
S.~Brown,
``The C4 model for visualising software architecture,''
\url{https://c4model.com/}, 2018.

\bibitem{openapi}
OpenAPI Initiative,
``OpenAPI Specification,''
\url{https://spec.openapis.org/oas/latest.html}, 2021.

\bibitem{yang2024sweagent}
J.~Yang et al.,
``SWE-agent: Agent-computer interfaces enable automated software engineering,''
in \emph{Proc.\ NeurIPS}, 2024.

\bibitem{zheng2023judging}
L.~Zheng et al.,
``Judging LLM-as-a-judge with MT-bench and Chatbot Arena,''
in \emph{Proc.\ NeurIPS}, 2023.

\bibitem{anthropic2026claude}
Anthropic,
``The Claude model family,''
\url{https://docs.anthropic.com/en/docs/about-claude/models}, 2026.

\bibitem{vasilevski2026beyondcorrectness}
K.~Vasilevski, X.~Dong, B.~Rombaut, R.~Deng, J.~Lin, A.~Leung, D.~Lin, B.~Chen, S.~Wang, and A.~E.~Hassan,
``Beyond correctness: Enhancing architectural reasoning in code LLMs via scalable labeling with agentic judgment,''
\emph{arXiv preprint arXiv:2606.14948}, 2026.

\bibitem{konrad2026architecture}
P.~M.~Konrad, T.~L.~Adam, R.~Terrenzi, and S.~Ayvaz,
``Architecture without architects: How AI coding agents shape software architecture,''
in \emph{Proc.\ IEEE ICSA}, 2026.

\bibitem{kammakomati2024concodeeval}
M.~Kammakomati, S.~Pimparkhede, S.~G.~Tamilselvam, P.~Kumar, and P.~Bhattacharyya,
``ConCodeEval: Evaluating large language models for code constraints in domain-specific languages,''
\emph{arXiv preprint arXiv:2407.03387}, 2024.

\bibitem{shafin2025process}
W.~I.~Shafin, M.~N.~Rafi, Z.~Li, and T.-H.~Chen,
``Evaluating software process models for multi-agent class-level code generation,''
\emph{arXiv preprint arXiv:2511.09794}, 2025.

\end{thebibliography}

\end{document}